\documentstyle[preprint,aps]{revtex}

\def\lll{\lambda}

\newcommand{\be}{\begin{equation}}
\newcommand{\ee}{\end{equation}}

\input epsf

\begin{document}
\preprint{WISC-MILW-96-TH-22}
\title{Choptuik scaling and Quantum effects in 2D Dilaton Gravity}
\author{Yoav Peleg,\footnote{Electronic address: {\em yoav@csd.uwm.edu}}
Sukanta Bose,\footnote{Electronic address: {\em bose@csd.uwm.edu}} and  
Leonard Parker\footnote{Electronic address: {\em leonard@cosmos.phys.uwm.edu}}  
}
\address{Department of Physics \\
University of Wisconsin-Milwaukee, P.O.Box 413 \\
Milwaukee, Wisconsin 53201, USA}
\date{\today}
\maketitle
\vskip 0.2cm
\begin{abstract}
We study numerically the collapse of massless scalar fields in two-dimensional 
dilaton gravity, both classically and semiclassically. 
At the classical level, we find that the black hole mass scales at threshold 
like $M_{\rm bh} \propto |p-p^*|^{\gamma}$, where $\gamma \simeq 0.53$. 
At the semiclassical level, we find that in general 
$M_{\rm bh}$ approaches a non-zero constant 
as $p \rightarrow p^*$. Thus, quantum effects produce a mass gap 
not present classically at the onset of black hole formation.
\end{abstract}

\newpage

The discovery of universality and scaling at the onset of black hole formation 
\cite{Choptuik} may have important implications in understanding the structure of 
solution-space in gravitational theory. 
Choptuik studied numerically the collapse of a spherically symmetric 
self-gravitating real scalar field in four-dimensional (4D) Einstein gravity 
and considered one-parameter 
families of initial data, ${\cal S}[p]$, where $p$ is a parameter specifying the 
strength of the gravitational self-interaction of the scalar field. 
He found that there exists a 
critical value, $p=p^*$, such that for $p<p^*$, the ``subcritical case'', 
no black hole is formed and the solution is regular, while  
for $p>p^*$, the ``supercritical case'', a black hole is formed. 
Furthermore, by careful analysis of the solutions near criticality, 
$p = p^*$, he found that as $p$ approaches $p^*$ 
from above, the mass of the created black hole approaches zero, 
and when a black hole just forms, its mass scales as 
$M_{\rm bh} \propto |p-p^*|^{\gamma}$, where the critical exponent 
$\gamma \simeq 0.37$. The critical solutions also exhibit discrete 
self-similarity \cite{Choptuik}. Similar behavior was found in other models 
of non-linear gravity \cite{Abrahams}. 

In the semiclassical scenario, i.e., for a quantum field propagating on 
a classical dynamical background metric, 
the created black hole of mass $M$ radiates in 4D 
with the Hawking temperature $T_{\rm H} \propto M^{-1}$ \cite{Hawking}. 
As $M \rightarrow 0$ near criticality, 
$T_{\rm H}$ becomes large and quantum effects are clearly important. 
To gain some understanding of how quantum effects may alter the classical 
scenario near criticality, we investigate a tractable two-dimensional (2D) model 
that exhibits classical scaling as well as significant quantum effects near 
criticality.

The model we study is 2D dilaton gravity coupled to a massless 
scalar field. We first consider the classical theory and then include quantum 
effects. We find that classically there is universal scaling of the black 
hole mass near criticality, $M_{\rm bh} \propto |p - p^*|^{\gamma}$, where 
$\gamma \simeq 0.53 \pm 0.01$ is independent of initial data. 
In addition, we find that the ground state of the classical theory gives 
a lower bound on the energy of spacetimes resulting from dynamical 
processes in which a black hole is formed. This is related to a radiation 
energy deficit, $\Delta_{\rm rad}$, that we describe later.  
At the semiclassical level, although the Hawking temperature 
in our 2D model is independent of $M$, we find that quantum 
effects alter significantly the critical behavior. 
Most interestingly, when $p$ is 
sufficiently close to $p^*$, the classical scaling behavior breaks down 
and $M_{\rm bh}$ approaches a non-zero constant value that depends on the 
initial data. The existence of this mass gap at the quantum level is a new 
result that may be of fundamental interest.

The classical theory is described by the 2D 
Callan-Giddings-Harvey-Strominger (CGHS) action \cite{CGHS} 
  \be \label{CGHS}
S_{\rm CGHS} = {1\over 2\pi} \int d^2 x \sqrt{-g} \biggl\{ e^{-2\phi} \Bigl[ 
{\cal R} +  4 (\nabla \phi)^2 + 4 \lll^2 \Bigr] - {1\over 2} \sum_{i=1}^{N} 
(\nabla f_i)^2 \biggr\} , 
  \ee
where $\phi$ is the dilaton field, ${\cal R}$ is the 2D Ricci scalar 
formed from the metric tensor $g_{\mu \nu}$, $\lll$ is 
a positive constant, and the $f_i$ are $N$ massless scalar matter fields 
conformally coupled to the 2D geometry. 
We work in the conformal gauge, 
$g_{--}=g_{++}=0$ and $g_{+-}=-(1/2)\exp(2\rho)$, (i.e., $ds^2 = 
- \exp[2\rho(x^+,x^-)] dx^+ dx^-$), 
where $(x^+,x^-)$ are the ``Kruskal'' null coordinates 
in which $\phi(x^+,x^-)=\rho(x^+,x^-)$ \cite{BPP2}. 

The general vacuum solution of the classical theory (\ref{CGHS}) is
$\phi = -(1/2)\ln(-\lll^2 x^+ x^- + C)$, where $\lll C$ is its ADM mass \cite{Witten}. 
For $C>0$ the vacuum solutions describe static 2D black holes. 
The $C=0$ solution is the linear dilaton vacuum (LDV), which is the ground state of the 
theory. Solutions with $C<0$ have timelike naked singularities in the strong coupling 
region where $\exp(2\phi) \rightarrow \infty$. 
The weak coupling, asymptotically flat region will be taken to be on the 
right-hand-side (rhs) of the spacetime \cite{BPP2}. 
In this work we take $C<0$, but we avoid the region 
of strong coupling and the singularity by considering the solutions only in the 
weak coupling region $\exp(2\phi) \leq \exp(2\phi_c)$ for a given constant $\phi_c$, 
and by imposing reflecting boundary 
conditions on the timelike boundary curve $\phi(x^+,x^-) = \phi_c$. 
A previous study of such a system at the onset of 
black hole formation \cite{StromThor} was based on boundary conditions  
that mix classical and one-loop contributions and do not have a standard 
classical limit. The reflecting boundary conditions \cite{PRL,Das} 
that we impose here have a standard classical limit. 
Let the boundary be described by the curve $x^+ = x^+_B(x^-)$. 
Then the reflecting boundary condition is 
  \be \label{boundaryequation}
T_{--}(x^-) = {\left( {\partial x^+_B \over \partial x^-} \right)}^2 
T_{++}(x^+_B(x^-)) \;, 
  \ee
where $T_{\pm \pm} = (1/2)\sum_i (\partial_{\pm} f_i)^2$ are the components, in the 
$(x^+,x^-)$ coordinates, of the stress tensor of the massless scalar fields. 
The general solution for the conformally coupled matter fields 
is $f_i(x^+,x^-) = f^+_i(x^+) + f^-_i(x^-)$. The initial data at right asymptotic past 
null infinity, $\Im^-_R$, are therefore given by $f^+_i(x^+)$ or $T_{++}(x^+)$. 
Working in the large $N$ limit \cite{BPP2} we are able to choose the 
boundary curve to be at $\exp(-2\phi_c) \rightarrow 0^+$ and get a second order 
non-linear ordinary differential equation (ODE) for the classical boundary curve 
\cite{BPP2}
  \be \label{ODE}
\left[ x^- + P_+(x^+_B)/{\lll}^2 \right] {d^2 x^+_B\over d{x^-}^2} + 
2{ dx^+_B\over dx^-} + 
2 \lll^{-2} T_{++}(x^+_B) {\left({dx^+_B\over dx^-}\right)}^2 = 0 \; , 
  \ee
where $P_+(x^+) = \int_0^{x^+} T_{++}(\bar{x}^+) d\bar{x}^+$. We solve the 
ODE (\ref{ODE}) numerically for different initial data, 
$T_{++}(x^+)$, with compact support, $x^+_1 < x^+ < x^+_2$. To the past of $x^+_1$, 
i.e., for $x^+<x^+_1$, we have a vacuum solution $\phi=-(1/2)\ln(-\lll^2 x^+ x^- + C)$, 
with $C$ being a negative constant. In this region we have an analytical solution for the 
boundary curve: $x^+_B(x^-) = C/x^-$, where $x^-<C/x^+_1$. 
For $x^- > C/x^+_1$, we integrate (\ref{ODE}) numerically to find the boundary curve 
and the corresponding solutions for the metric and dilaton field. 

We study in detail the two families of initial data shown in Fig.\ \ref{fig1}. 
Family ($\alpha$) is a two-parameter family specified 
by the amplitude, $A$, and the inverse width, $\Gamma$, of the profile $T_{vv}(v)$, 
where $v=\lll^{-1} \ln(\lll x^+)$ and $u=\lll^{-1} \ln(-\lll x^- - P_+(\infty)/\lll)$ 
are the manifestly asymptotically flat null coordinates on $\Im^{\pm}_R$. 
Family ($\beta$) is a 
five-parameter family described by $A_1$, $A_2$, $\Gamma_1$, $\Gamma_2$ and $\delta$. 
All the parameters are scaled by appropriate powers of $\lll$ to make them 
dimensionless. In these two cases (and in other cases that we studied) 
we find that fixing all but one arbitrarily chosen parameter, say $p$, 
yields a regular evolution with no black hole formation {\em if and only if} $p<p^*$. 
The critical value, $p^*$, depends on the values of the 
other fixed parameters. As in the 4D solution-space, the only ``intermediate solution'' 
separating black hole solutions from regular ones  
is the critical solution with $p=p^*$. 

Next we show that there is a universal scaling of the black hole mass near 
criticality in the classical theory. 
In order to find the mass of the created black hole 
we calculate the outgoing radiation reaching right asymptotic future null infinity, 
$\Im^+_R$, after being reflected from the boundary. 
This radiation is described by the stress tensor, $T_{uu}(u)$. 
Then we find that the Bondi mass at late times ($u \rightarrow \infty$) on $\Im^+_R$ 
is $M^{(\infty)}_{\rm Bondi} \equiv 
M_{\rm Bondi}(u \rightarrow \infty) = M_{\rm ADM} 
- \int_{-\infty}^{\infty} T_{uu}(u)du$. 
Here $M_{\rm ADM}$ is the ADM mass \cite{Billal} of the spacetime, 
defined such that the LDV ground state has zero ADM mass. 
In Fig.\ \ref{fig2} we show the 
numerical results for the reflected outgoing radiation, $T_{uu}(u)$, for two nearby 
sets of initial data. The two solutions correspond to family ($\alpha$) 
with the same width, $\Gamma = 0.2$, but with different amplitudes, $A$. 
The upper gray curve corresponds to a subcritical solution near criticality, 
$A = A_{\rm sub} = A^* - \epsilon_1 = 
1.546$. The black curve corresponds to a supercritical case near criticality, 
$A = A_{\rm super} = A^* + \epsilon_2 = 1.547$, where $A^*$ is the critical value 
of the amplitude parameter $A$.  

Although the 
initial data are very similar in the two cases, the late-time outgoing 
radiation is quite different. For the subcritical solution near criticality, the 
boundary curve becomes almost null and the late-time outgoing radiation 
is highly blue-shifted \cite{BPP2}. 
In the supercritical case on the other hand, this blue-shifted 
radiation is absent, and we find that part of the incoming radiation does not reach 
future asymptotic null infinity. Since $T_{uu}$ is quite different in the two cases, 
the Bondi mass defined through the integral  
$\int T_{uu}(u) du$ is obviously different. 
For all the subcritical solutions 
the Bondi mass at late times, $M_{\rm Bondi}^{(\infty)}$, is equal to 
the negative ADM mass of the initial spacetime, $\lll C$. 
However, for the supercritical solutions 
$M_{\rm Bondi}^{(\infty)}$ is always positive. We define the 
``radiation energy deficit'', $\Delta_{\rm rad}$, to be $\Delta_{\rm rad} \equiv 
\mbox{lim}_{\epsilon \rightarrow 0} \left\{ M^{(\infty)}_{\rm Bondi}[p=p^*+\epsilon] - 
M^{(\infty)}_{\rm Bondi}[p=p^*-\epsilon] \right\}$. 
The fact that $\Delta_{\rm rad}$ turns out to be  
non-zero is a striking manifestation of the non-linearity of our problem. 

We find that $\Delta_{\rm rad}$ does not depend on the profile of the 
infalling matter, but only on the initial vacuum geometry. 
Explicitly, 
  \be \label{Deltamass}
\Delta_{\rm rad} = - \lll C = \lll |C| \; , 
  \ee
where $C$ is the constant specifying the initial vacuum geometry. 
The existence of this non-zero $\Delta_{\rm rad}$ can be interpreted as 
implying that in the classical theory the LDV is a ``stable ground state''. 
In particular, we have started with a negative-mass geometry and find that 
the process of throwing in matter to form a black hole results in spacetimes 
having non-negative mass relative to the LDV. In agreement with 
cosmic censorship, this suggests that systems 
in this classical theory will not evolve to states of energy lower than the 
LDV, which have naked singularities. 
Once the infalling matter configuration is dense enough, $p>p^*$, 
part of the incoming energy is ``swallowed'' by the strong curvature region 
to compensate for the negative mass of the initial spacetime and  
make the future spacetime a solution with positive energy compared 
to the LDV. For our initial geometry with negative ADM mass
$\lll C$, the minimum energy that 
must be swallowed in order to get a positive-energy spacetime, is precisely 
$\Delta_{\rm rad}$. The resulting positive-energy spacetime is a 2D black hole. 
We find that in this classical theory 
the mass of the created black hole, $M_{\rm bh} \equiv 
M_{\rm Bondi}^{(\infty)}[p>p^*]$, approaches zero as $p$ approaches $p^*$ from above. 
Moreover, the black hole mass scales as 
  \be \label{bhmass}
M_{\rm bh} \propto |p-p^*|^{\gamma}  
  \ee
at threshold, where $\gamma = 0.53 \pm 0.01$. The value of $\gamma$ 
does not appear to depend on the profile of the infalling matter or the 
initial geometry. In Fig 3(a) we plot $\ln(M_{\rm bh}/\lll) + a_i$ 
versus $\ln|p_i-p_i^*|$ for different parameters $p_i$. The constants 
$a_i$ are chosen such that the three curves intersect at a given point. 
The slopes of all the lines are the same within our numerical accuracy. 

Next we consider quantum effects. We quantize the scalar fields on a classical 
dynamical background geometry consisting of the metric and the dilaton field. 
The semiclassical effective action that we study is \cite{BPP1}
  \begin{eqnarray} \label{semi}
S_{\rm eff} &=& S_{\rm CGHS} - {\kappa \over 8 \pi} \int d^2x \sqrt{-g(x)} 
\int d^2x' \sqrt{-g(x')} {\cal R}(x) G(x,x') {\cal R}(x') \nonumber \\
        & & + {\kappa \over 2 \pi} \int d^2 x \left[ (\nabla \phi)^2 - \phi {\cal R} 
\right] \; , 
  \end{eqnarray}
where $\kappa = N \hbar / 12$ and $G(x,x')$ is an appropriate Green function for the 
massless scalar fields. 
The second term on the rhs of Eq. (\ref{semi}) is the 
Polyakov-Liouville effective action derived from the trace anomaly of the 2D massless 
scalar fields \cite{Polyakov}, and the last term on the rhs of Eq. (\ref{semi}) 
is a local counter-term that we add to our effective theory in order to make it exactly 
solvable \cite{BPP1}. The vacuum solutions of our semiclassical theory are 
$\phi = -(1/2) \ln[-\lll^2 x^+ x^- - (\kappa/4)\ln(-\lll^2 x^+ x^-) + \bar{C}]$, 
where $\bar{C}$ is a constant. The ground state is the solution with 
$\bar{C} = \bar{C}_0 \equiv (\kappa / 4)[\ln(\kappa/4) - 1]$. 
As in the classical case, we impose reflecting boundary conditions on 
a timelike boundary curve and study the solutions only in the weak coupling region. 
The reflecting boundary condition is a modification of Eq. 
(\ref{boundaryequation}) where $T_{\pm \pm}$ are now components of the total 
stress tensor, including the classical and one-loop contributions. There is also 
an extra term due to possible particle creation from the boundary, which is effectively 
a moving mirror \cite{Birrell}. One can eliminate the moving mirror term by 
taking the large $N$ limit together with $\exp(-2\phi_{c}) \rightarrow 0^+$ \cite{BPP2}. 
In contrast to (\ref{ODE}), the resulting nonlinear ODE 
for the semiclassical boundary curve, $x^+=x^+_B(x^-)$, is 
\cite{BPP2} 
  \be \label{quantumboundary}
\left[ x^- + {P_+(x^+_B)\over \lll^2} + {\kappa\over 4 \lll^2 x^+_B} \right] 
{d^2 x^+_B\over d{x^-}^2} + 2{ dx^+_B\over dx^-} +
\left[ 2 \lll^{-2} T_{++}(x^+_B) - {\kappa\over 4 \lll^2 {x^+_B}^2} \right] 
{\left({dx^+_B\over dx^-}\right)}^2  = 0  . 
  \ee

We consider initial data with no quantum radiation on $\Im^-_R$ and classical 
incoming radiation corresponding to the profiles shown in Fig.\ \ref{fig1}. 
We find that as in the classical case, the solutions are regular and no black hole 
is formed if and only if $p < p^*_q$ (where 
$p^*_q$ is the critical value of the parameter $p$ in the semiclassical 
case). For $p>p^*_q$, the black hole that is formed 
evaporates by emitting Hawking radiation \cite{BPP1}. 
To look for scaling analogous to Eq. (\ref{bhmass}), 
we calculate the {\em initial} mass of the created black hole. 
It is this mass that reduces in the limit $N \hbar \rightarrow 0$ 
to the mass of the classical black hole that appears in Eq. (\ref{bhmass}). 
The initial black hole mass, 
$M^{(i)}_{\rm bh}$, is the Bondi mass at $u=u_0$, where $u_0$ is the minimum of the 
apparent horizon $u=u_{\rm ah}(v)$. The apparent horizon 
is the solution of the equation $\partial_{+} \phi = 0$. 
Explicitly, $M_{\rm bh}^{(i)} = M_{\rm ADM} - \int_{-\infty}^{u_0} T_{uu} du$, 
where $T_{uu}(u)$ is the total stress tensor of the outgoing radiation and 
$M_{\rm ADM}$ is the ADM mass of the spacetime, defined such that 
the semiclassical ground state, i.e., the static vacuum solution with 
$\bar{C}=\bar{C}_0$, has zero ADM mass. 

We examine quantum effects by considering cases with different values 
of $\kappa = N \hbar / 12$. 
In Fig.\ \ref{fig3}(b) we plot $\ln(M_{\rm bh}^{(i)}/\lll)$ versus 
$\ln|p-p^*_q|$ for two different values of $\kappa$. We do so in the case 
of the family ($\alpha$) of initial data shown in Fig.\ \ref{fig1}, 
where the free parameter $p$ is the amplitude $A$, 
and the parameter $\Gamma = 0.2$ is fixed. 
For large values of $M^{(i)}_{\rm bh}$ the initial mass of the 
black hole behaves like that of the classical black hole 
(\ref{bhmass}). However as $M^{(i)}_{\rm bh}$ decreases, deviations from the 
classical behavior appear. Unlike the classical case, 
as $p$ approaches $p^*_q$ the initial mass, $M^{(i)}_{\rm bh}$ of the 
created black hole {\em does not} generally approach zero, but rather 
approaches a constant, $M_{\rm gap}$. 
We find that this mass gap, $M_{\rm gap}$, 
depends not only on the value of $\kappa$ but also on the initial data. 
For the cases shown in Fig.\ \ref{fig3}(b) we have 
$M_{\rm gap}/\lll \simeq 0.014$ for $\kappa = 10^{-2}$, 
and $M_{\rm gap}/\lll \simeq 0.0037$ for $\kappa = 10^{-3}$. 
The radiation energy deficit in the quantum case 
is $\Delta^{q}_{\rm rad} = \lll(\bar{C}_0 - \bar{C}) + M_{\rm gap}$. 
 
The deviations from the classical scaling law (\ref{bhmass}) become significant 
for values of $M^{(i)}_{\rm bh}$ that are of order $\kappa \lll$. When 
$M^{(i)}_{\rm bh}$ takes values that are of order or less than $\kappa \lll$,  
the semiclassical approximation remains valid provided $N$ is sufficiently large 
\cite{BPP2,BPP3}.
However for any fixed value of $N$, no matter how large, the semiclassical 
approximation breaks down when $p$ is sufficiently close to $p_q^*$
\cite{BPP2}, i.e., as 
one moves to the far left well beyond the region shown in Fig.\ \ref{fig3}(b). 
Thus, determining the behavior of the curve as $\ln|p-p^*_q| \rightarrow - \infty$ 
requires full quantization of the theory. We are currently studying the phase 
structure of the semiclassical theory near threshold which appears to be richer 
than that in the classical case.
      
In this work we investigate numerically the collapse 
of massless scalar fields in 2D dilaton gravity. We find that classically the 
black hole mass at threshold obeys a power-law, $M_{\rm bh} \propto |p-p^*|^{\gamma}$, 
where $\gamma = 0.53 \pm 0.01$. However, quantum effects give rise to a mass gap 
that depends on the initial data.

\vspace{0.5cm}

We would like to thank B. Allen, J. Friedman, D. Garfinkle, T. Piran, and A. Steif 
for very helpful discussions. This work was supported by the National Science 
Foundation under Grant No. PHY 95-07740.

\newpage

  \begin{figure} 
\caption{Two families of initial data described by the stress tensor $T_{vv}$.}
\label{fig1}
  \end{figure}
  \begin{figure} 
\caption{The stress tensor $T_{uu}$, in units of $\lll^2$, 
describing the outgoing radiation. The upper gray curve 
corresponds to a subcritical solution just below criticality and the black curve 
corresponds to a supercritical solution just above criticality.}
\label{fig2}
  \end{figure}
  \begin{figure} 
\caption{$\ln(M_{\rm bh}/\lll)$ versus $\ln|p-p^*|$ in the classical case, (a), and the 
quantum case, (b).}
\label{fig3}
  \end{figure}

\end{document}